\documentclass[fleqn, usenatbib]{mnras}

\usepackage{newtxtext}
\usepackage{newtxmath}

\usepackage[T1]{fontenc}
\usepackage{pdflscape}
\usepackage{longtable}
\usepackage{booktabs}
\usepackage{color}
\usepackage{amssymb}
\usepackage{mathtools}
\usepackage{xspace}
\usepackage{rotating}
\usepackage{appendix}
\usepackage{multirow}
\usepackage{multicol}
\usepackage{array}
\usepackage{subcaption}
\usepackage{comment}
\usepackage{tabularx}
\usepackage{bigstrut}
\usepackage{threeparttable}
\usepackage{afterpage}
\usepackage{capt-of}
\usepackage{newtxtext,newtxmath}
\usepackage[T1]{fontenc}
\usepackage{ae,aecompl}
\usepackage{graphicx}	
\usepackage{amsmath}	
\usepackage{amssymb}
\usepackage{blindtext}
\usepackage{comment}
\usepackage{tabularx}
\usepackage{tabulary}
\usepackage{bigstrut}
\usepackage{threeparttable}
\usepackage{afterpage}
\usepackage{capt-of}
\usepackage{caption}
\usepackage{subcaption}
\usepackage{hyperref}

   \newcommand{\ecyc}{E$_{\texttt{cyc}}$} 
   
   \newcommand{\Lc}{$L_{\texttt{crit}}$}
    \newcommand{\astro}{\textit{AstroSat}}
    \newcommand{\hxmt}{{\textit{Insight}}-HXMT}
    \newcommand{\nustar}{\textit{NuSTAR}}
    \newcommand{\beppo}{\textit{BeppoSAX}}
    
\title[\textit{AstroSat} and \hxmt\ view of  XTE J1946+274]{\textit{Insights} into the phase-dependent cyclotron line feature in XTE J1946+274: An \textit{AstroSat} and \textit{Insight}-HXMT view}
\author[Devaraj et al.]{Ashwin Devaraj,$^{1,2}$\thanks{\href{mailto:ashwin@rri.res.in}{ashwin@rri.res.in}}  
Rahul Sharma,$^1$
Shwetha Nagesh,$^{1,3}$
Biswajit Paul$^{1}$\thanks{ \href{mailto:bpaul@rri.res.in}
{bpaul@rri.res.in}}\\
$^{1}$Raman Research Institute, Sadashivanagar, Bangalore-560080, India\\
$^{2}$Joint Astronomy Programme, Indian Institute of Science, Bangalore-560012, India\\
$^{3}$Department of Physics \& Astronomy, Texas Tech University, Box 41051, Lubbock, TX, 79409-1051, USA\\}

\begin{document}
\label{firstpage}
\pagerange{\pageref{firstpage}--\pageref{lastpage}}

\maketitle
\begin{abstract}
XTE J1946+274 is a Be/X-ray binary with a 15.8s spin period and 172 d orbital period. Using \textit{RXTE/PCA} data of the 1998 outburst, a cyclotron line around 37 keV was reported. The presence of this line, its dependence on the pulse phase, and its variation with luminosity have been of some debate since. In this work, we present the reanalysis of two \astro \ observations: one made during the rising phase of the 2018 outburst and the other during the declining phase of the 2021 outburst. We also present a new analysis of the \textit{Insight}-HXMT observations of the source at the peak of the 2018 outburst. We find the source to be spinning up over the course of the outburst and spinning down between the two outbursts. We report the presence of a higher cyclotron line energy using the 2018 \astro \ observation ($\sim 45$ keV) and 2018 \hxmt \ observation ($\sim$ 50 keV) and a line at $\sim$ 40 keV during the declining phase of the 2021 outburst using data from \astro. We also investigate the pulse phase dependence of the cyclotron line parameters and find that the line is significantly detected in all the phases of both \astro \ observations, along with showing variation with the pulse phase. This differs from the previous results reported using \beppo \ and \nustar. We explain this behaviour of the cyclotron line to be due to photon spawning and different accretion column radii at the two poles of this neutron star.

\end{abstract}

\section{INTRODUCTION}
Cyclotron Resonant Scattering Features (CRSF) detected in the hard X-ray spectrum of some high-mass X-ray Binaries (HMXB) are the best diagnostic tools to study the magnetic field strength of these neutron star systems. CRSF or Cyclotron lines are absorption-like features formed as a result of scattering of the hard X-ray photons from the accretion column of the neutron star at resonant energies of the electrons in the magnetic field due to the magnetic field strength of the neutron star. The relation between the cyclotron line centroid energy and the magnetic field strength can be approximately represented by the equation E$_{\mathrm{cyc}} \sim 11.6 \  n \ \mathrm{B}_{12}$ keV, where B$_{12}$ is the magnetic field strength in the units of 10$^{12}$ G and $n$ represents the harmonic. (See \citealt{Staubert_2019} for a review on cyclotron lines in neutron stars.)  

Out of over 150 HMXBs that have been discovered in the Galaxy \citep{Fortin_2023}, only a third of the sources have been confirmed to exhibit a cyclotron line feature in their spectrum \citep{Staubert_2019}. Cyclotron line parameters are known to vary depending on the pulse phase \citep{staubert_2014,Varun_2019a}, luminosity \citep{Tsygankov_2006, Rothschild_2017, Vasco_2011} and time \citep{staubert_2014, Bala_2020}. Several of these sources are also transients and have been observed only once or a few times.  

Discovered in September 1998 during a major outburst, XTE J1946+274 is a Be/X-ray binary (BeXRB) with a 15.8 s spin period, a $\sim170$ d orbital period \citep{Smith1998,Campana_1999} and a B0-1 V-IVe star for a companion \citep{Verrecchia_2002}. Using data from \texttt{IXAE}, \citet{Paul_2001} determined the structure of the pulse profiles of XTE J1946+274 to have a double-peaked structure. They also found the source to be spinning up over the course of the outburst. \citet{Wilson2003} found evidence for the presence of an accretion disk. They also determined the distance to the source to be around 9.5$\pm$2.9 kpc. Using \textit{RXTE/}PCA observations of this outburst, \citet{Heindl2001} reported the presence of a cyclotron line at $\sim 36.5$ keV, which was also reported by \citet{Doroshenko_2017} using archival \textit{BeppoSAX} observations of the 1998 outburst. During the next outburst of 2010, it was observed with \textit{RXTE} \citep{Muller2012} and \textit{Suzaku} \citep{Maitra2013, Marcu_Cheatham_2015}. \citet{Muller2012} reported a line at 25 keV, however, this was different from the 38 keV cyclotron line reported by \citet{Maitra2013}, and no subsequent evidence was found for a line at 25 keV. 

\begin{figure*}
    \centering
 \vspace{0.5cm}
	\includegraphics[scale=0.39,trim={0 1.0cm 0 2.0cm},  angle=0]{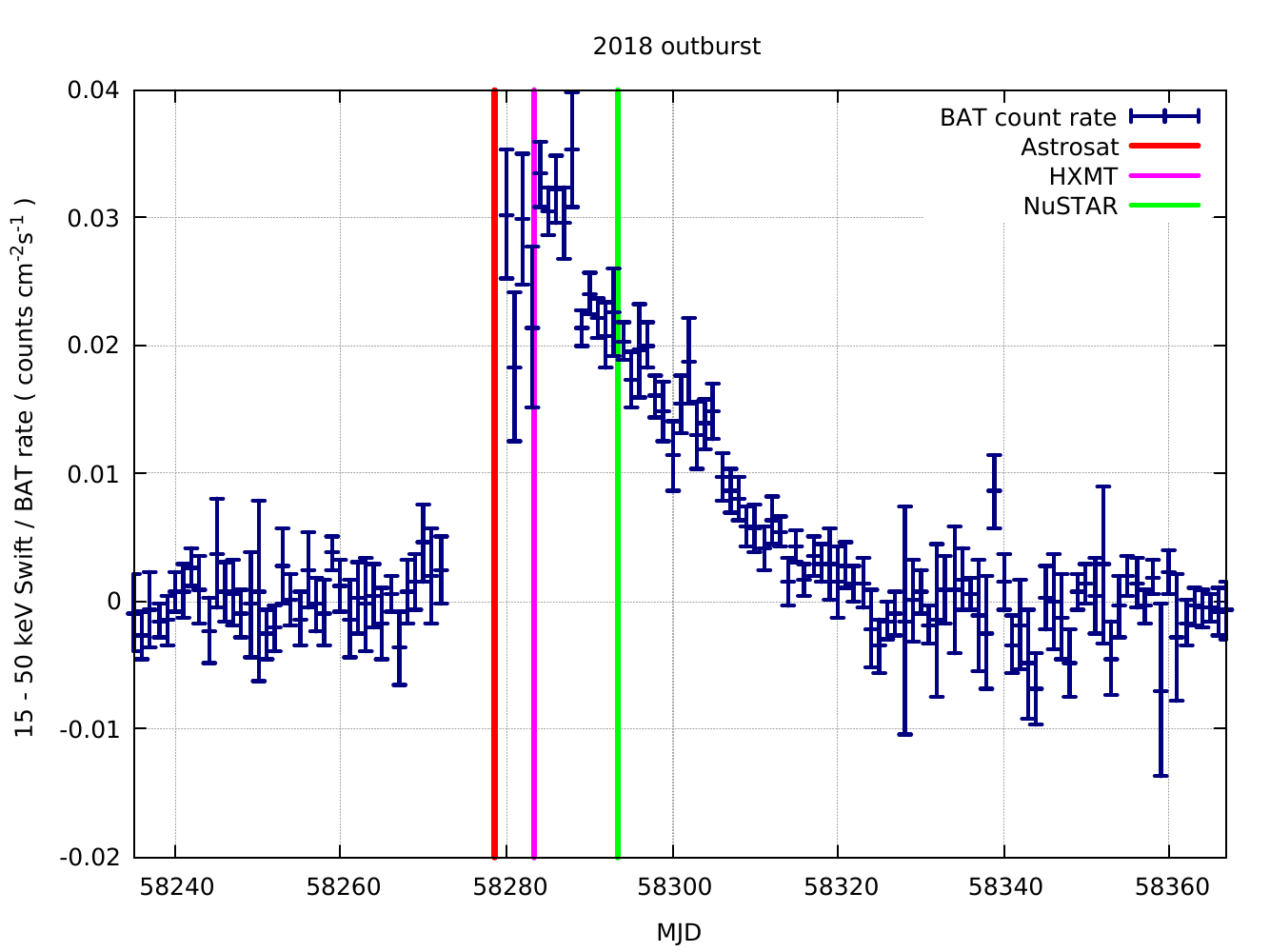}
     \includegraphics[scale=0.39,trim={0 1.0cm 0 2.0cm},  angle=0]{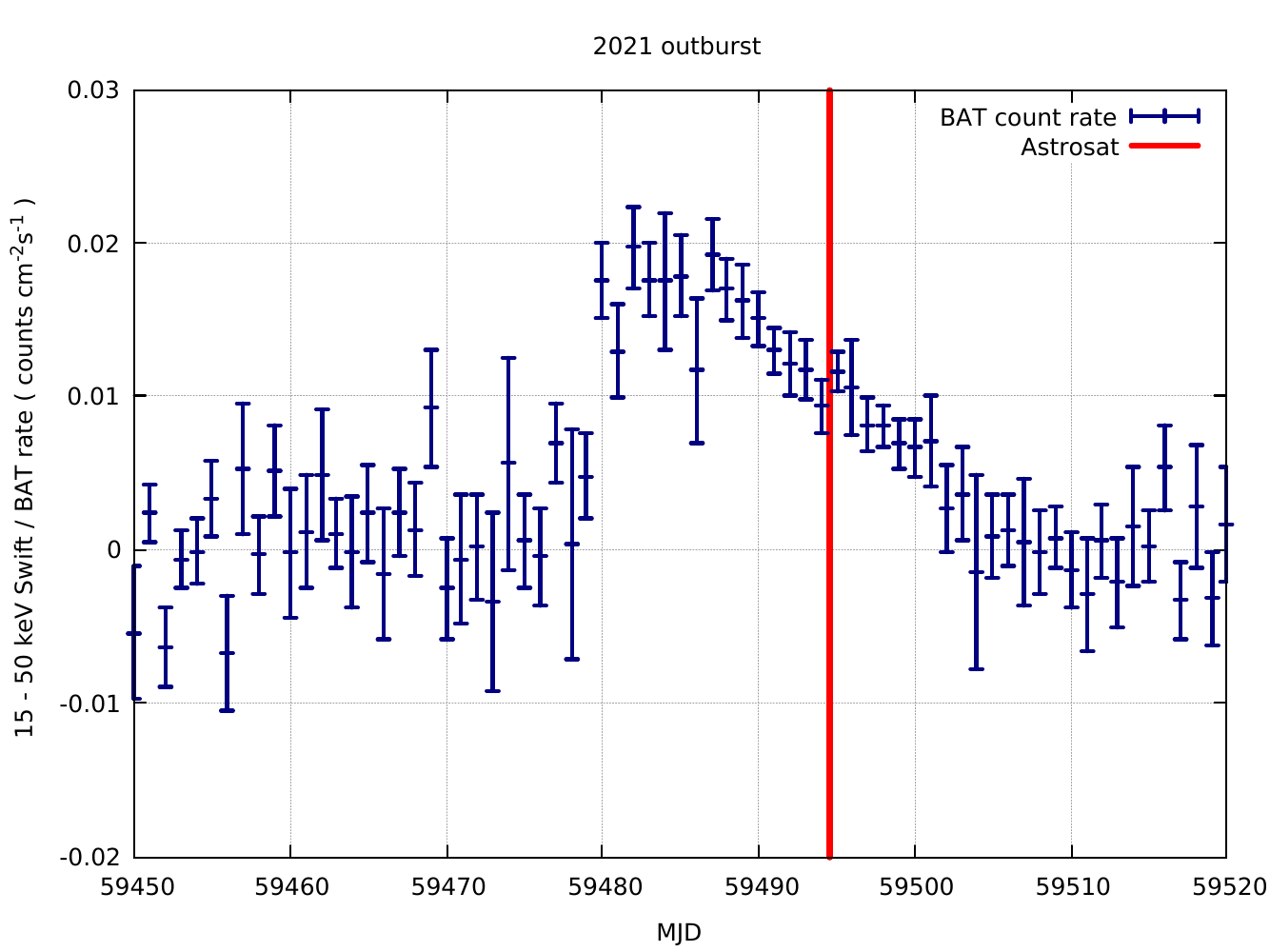}
    \caption{\textit{Swift/BAT} light curve of the 2018 outburst (left) and the \textit{Swift/BAT} light curve of the 2021 outburst (right). Blue points indicate the \textit{Swift/BAT} count rate while the red, magenta, and green vertical lines indicate the times of the \textit{AstroSat}, \textit{HXMT} and \textit{NuSTAR} observations, respectively. The data were obtained from \url{https://swift.gsfc.nasa.gov/results/transients/weak/XTEJ1946p274/}}
    \label{fig:outburst-lightcurve1}
\end{figure*}

In 2018, the source went into an outburst that lasted approximately 50 days and was observed with \astro, \hxmt\ and \nustar. The \nustar\ observation was made during the declining phase of the outburst (see left panel of Fig.\ref{fig:outburst-lightcurve1}) when the \textit{Swift BAT} rate was $\sim 0.02$ counts cm$^{-2}$s$^{-1}$ ($\sim91$ mCrab). Using this observation, a line at $\sim$ 38 keV was reported by \citet{Gorban_2021} and \citet{Devaraj_2022}. The cyclotron line centroid energy was found to be non-varying with pulse phase or luminosity \citep{Devaraj_2022}. The \astro\  observation of 2018 was made during the rising phase of the outburst, and the \hxmt\  observation was performed during the peak of the outburst where the \textit{Swift BAT} count rate reached almost 0.03 counts cm$^{-2}$s$^{-1}$($\sim136$ mCrab). The source underwent an outburst of a lower peak flux in 2021 with a maximum \textit{BAT} count rate of $\sim 0.02$ counts cm$^{-2}$s$^{-1}$ (see Fig. \ref{fig:outburst-lightcurve1}) and this was observed only with \astro \ during the declining phase when the flux reached half the peak luminosity($\sim45$ mCrab). Both the \astro \ observations were analyzed by \citet{chandra_2023} and reported a higher line energy at around 43 keV. \citet{chandra_2023} discuss the spin evolution over time and the possible reasons for the unusual outbursts of XTE J1946+274. 

The optical depth (and detectability) of the CRSF has been found to be highly dependent on the pulse phase of this source from results using \nustar \ \citep{Devaraj_2022} and \beppo \ \citep{Doroshenko_2017}.  In this work, we reanalyse the \astro \ 2018 and 2021 observations and the 2018 \hxmt \ observation to investigate the same at different luminosities. Different from \citet{chandra_2023}, we also confirm the presence of a 10 keV feature which has been reported in the spectrum of several X-ray pulsars, including XTE J1946+274 \citep{Coburn_2002,hemanth_2023}.

\section{OBSERVATION AND DATA REDUCTION}

The Hard X-ray Modulation Telescope named \textit{Insight}-HXMT was launched on 2017 June 15 in a low earth orbit with an altitude of 550 km at an inclination angle of 43 degrees \citep{zhang_2014,zhang_2020}. HXMT carries three instruments on board: the High Energy X-ray telescope (HE) uses 18 NaI(Tl)/CsI(Na) scintillation detectors and is sensitive to X-rays in the 20--250 keV band. It has a total geometrical area of about 100 cm$^2$ and the energy resolution is $\le$17\%@60 keV, with a time resolution of 25 $\mu$s \citep{Liu_2020}; the Medium Energy X-ray telescope (ME) consists of 1728 SiPIN detectors to detect photons in the 5--30 keV band using a total geometrical area of 952 cm$^2$. It has a time resolution of 280 $\mu$s and an energy resolution of 15\%@20 keV \citep{cao_2020}; the Low Energy X-ray detector (LE) contains 96 SCD detectors suitable for photons with energies in the range 1--15 keV and a geometrical area of 384 cm$^2$, with a time resolution of 1 ms and an energy resolution of 2.5\%@6 keV \citep{chen_2020}. The three payloads are co-aligned and simultaneously observe the same source.

We analyzed the observation P0114760001-P0114760013 consisting of 13 snapshots, during the peak of the 2018 outburst of the source from MJD 58283.32 to MJD 58284.94. The data were then screened using good time intervals created with the following criteria: the Earth elevation angle greater than 10 degrees, the cutoff rigidity greater than 8 GeV, and the offset angle from the pointing source less than 0.04 degrees. We also exclude the photons collected 300s before entering and after exiting the South Atlantic Anomaly region. The data reduction was carried out using The \textit{Insight}-HXMT Data Analysis Software Package (HXMTDAS) V2.05 and the corresponding Calibration Database (CALDB) v2.06. The data had exposures of 13.1 ks, 33.4 ks and 26.5 ks for the LE, ME, and HE detectors respectively. The photon arrival times were barycenter corrected using the \hxmt \ tool \texttt{hxbary2}.

Large Area X-ray Proportional Counter (LAXPC) is an instrument onboard the \astro\ satellite launched in 2015 \citep{agarwal_2006}. It houses 3 identical, co-aligned proportional counters (LAXPC10, LAXPC20, and LAXPC30) that operate in the 3-80 keV energy range independently recording the photon arrival times at a resolution of 10 $\mu$s. Due to low gain in LAXPC10 and LAXPC30 being switched off, we only use the data from LAXPC20 in this work. 

\astro \ observed the 2018 outburst between the $9^{\mathrm{th}}$ and $10^{\mathrm{th}}$ June 2018 (OBS ID 90000021480) with an exposure of $\sim 53$ ks. The second \astro \ observation was made between $3^{\mathrm{rd}}$ and $6^{\mathrm{th}}$ October 2021 with an exposure of $\sim 114$ ks (OBS ID: 9000004716).  We use the latest version of the LAXPC software package (August 2022 release) with improved background estimation accounting for the diurnal variation \citep{Antia_2022} for the extraction and reduction of the data\footnote{\url{http://astrosat-ssc.iucaa.in/laxpcData}}. The event files were barycenter corrected using the tool \texttt{As1bary}. 

Orbital corrections were done on the event files of all three observations using the updated ephemeris reported in the Fermi/GBM website\footnote{\url{https://gammaray.nsstc.nasa.gov/gbm/science/pulsars/lightcurves/xtej1946.html}} based on the orbital solutions from \citet{Wilson2003}.

\section{Analysis and Results}
\subsection{Timing Analysis}
The lightcurves were extracted at a resolution of 0.01 s and were then background subtracted for all the observations. For the \astro\ observation, we used only the top layer for the timing analysis. The periods for the 2018 \astro, the 2018 \hxmt \ and the 2021 \astro \ observations were determined to be P$_{\mathrm{as1}}$=15.75505(6) s, P$_{\mathrm{hx}}$= 15.75409(3) s and P$_{\mathrm{as2}}$=15.755143(4) s respectively. The period from \nustar \ observation, chronologically after the \hxmt \ observation, was found to be 15.75199 s \citep{Devaraj_2022}. This is indicative of a spin-up trend across the outburst due to the accretion torque applied on the neutron star \citep{Ghosh_1978}. In quiescence, as also noted by \citet{chandra_2023}, the source continues to spin down as can be inferred from the higher period estimated from the 2021 observation. Due to the long duration of the 2021 observation, which could result in the decoherence of the pulse shape if the change in frequency were not accounted for, we determined the period derivative for this observation to be $\dot {P} = -1.98(3) \times 10^{-9}\mathrm{s s}^{-1}$ using the same method as described in \citet{Devaraj_2022}.


\begin{figure*}
\centering
\includegraphics[scale=0.38,trim={1.0cm 1cm 0cm 0cm},clip ]{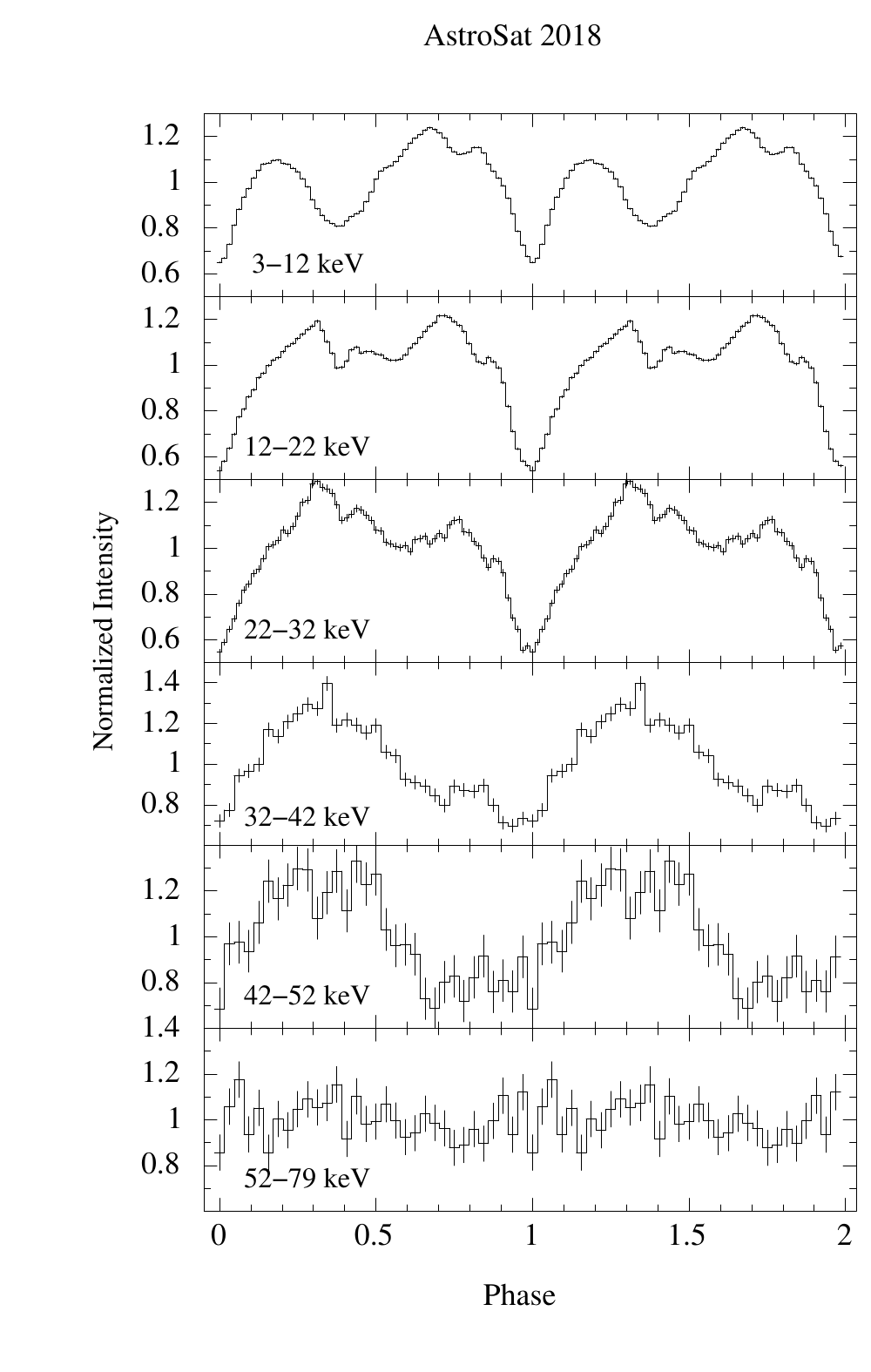} 
\includegraphics[scale=0.38,trim={1.0cm 1cm 0cm 0.0cm},clip ]{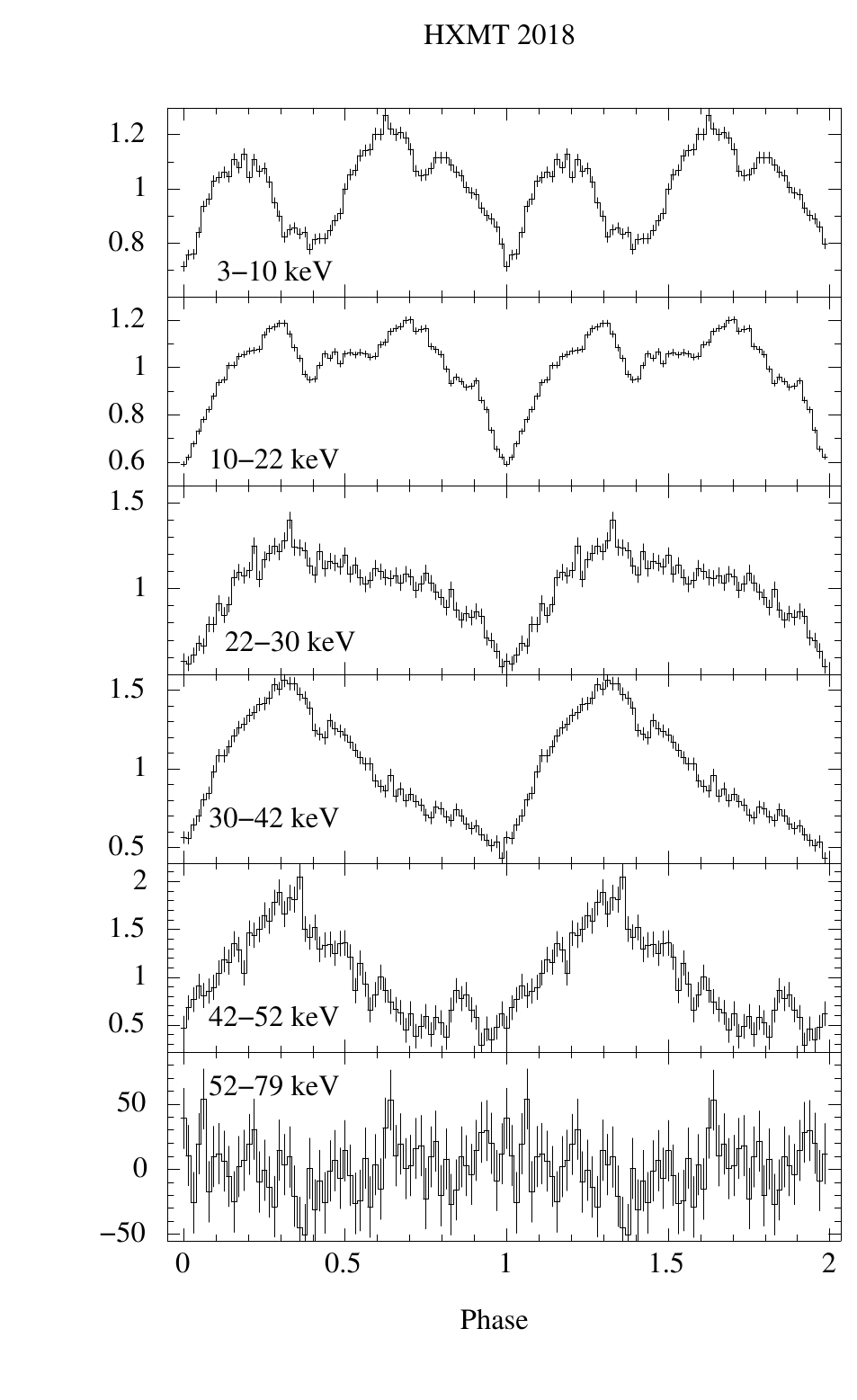}
\includegraphics[scale=0.38,trim={1.0cm 1cm 0cm 0cm},clip ]{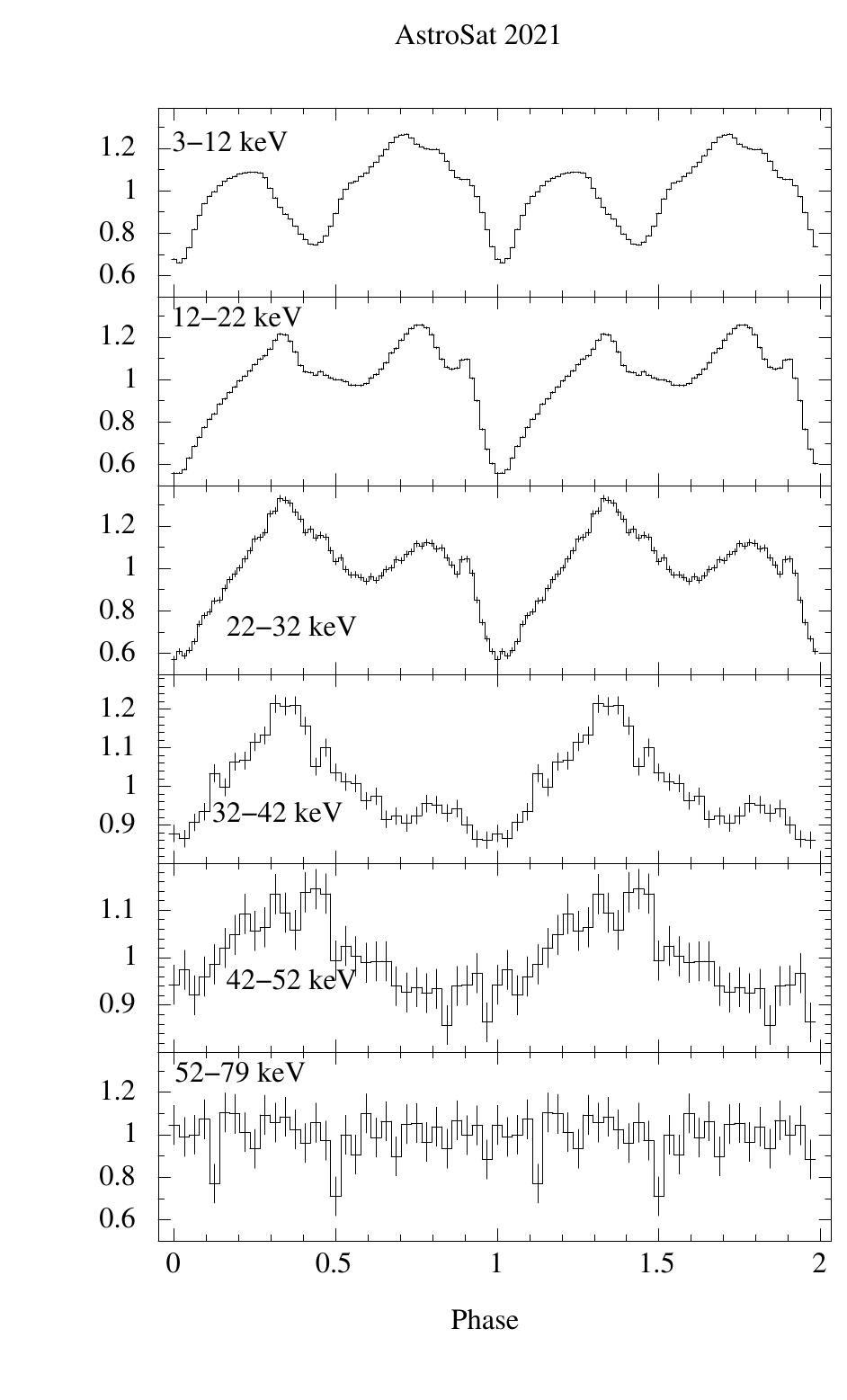}
\caption{Energy resolved pulse profiles of the \astro\ 2018 (left penal), HXMT 2018 (middle panel) and \astro\ 2021 (right panel) observations.  }
\label{fig:BBpulseprofiles-all_ast1}
\end{figure*}




\subsubsection{Energy-resolved pulse profiles}

We generated the energy-resolved pulse profiles in the 3--12 keV, 12--22 keV, 22--32 keV, 32--42 keV, 42--52 keV and 52--79 keV energy ranges for the \astro \ observations and folded them with the appropriate period and period derivative. For the \hxmt \ lightcurves we produced the energy-resolved pulse profiles in the 3-10 keV, 10-22 keV, 22-30 keV, 30-42 keV, 42-52 keV and 52-79 keV energy ranges.  From Fig. \ref{fig:BBpulseprofiles-all_ast1} and the \nustar\ pulse profiles reported in \citet{Devaraj_2022}, the pulse profiles exhibiting a double-peaked structure at lower energy ranges while evolving to a single-peaked structure at high energy ranges are common over the course of the 2018 outburst as well across outbursts. Pulsations are not detected above 52 keV. Apart from changes in a few micro-structures in the pulse profiles, the overall shapes are very similar in the same energy bands despite the source being observed at a range of fluxes.

\subsection{Broadband spectral analysis}

\begin{figure*}
    \centering
	\includegraphics[scale=0.65, trim={0 3.0cm 0 1.8cm},  angle=-90]{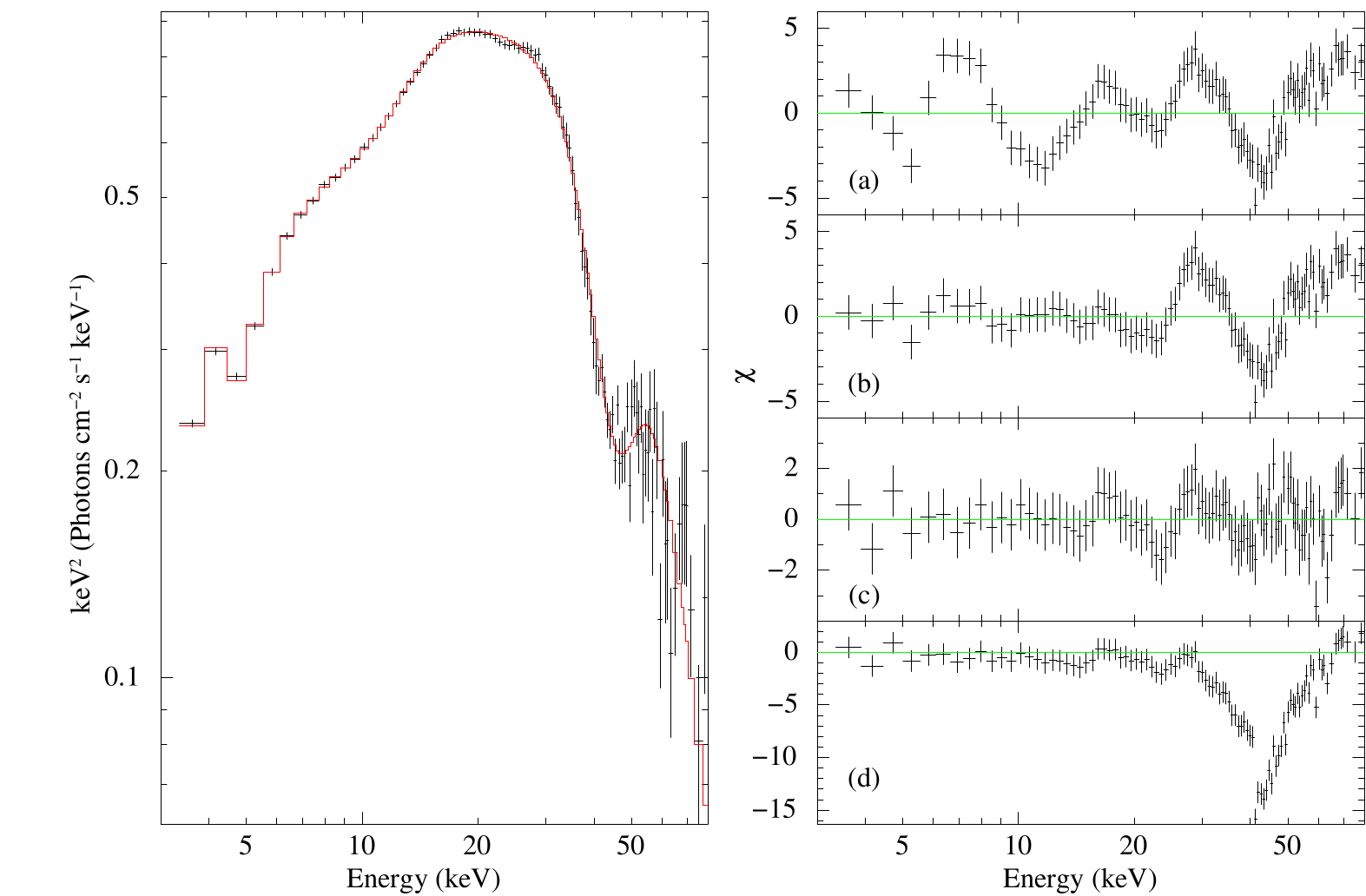}
    \caption{The unfolded spectrum of the \astro\ 2018 observation with the best-fit of \texttt{NPEX} model is shown on the left panel. The right panel shows the residues. Panel (a) is the fit without the cyclotron line at 45 keV and the 10 keV feature (b) is the fit including a \texttt{gabs} at 10 keV (c) is the best fit including a \texttt{gabs} at 45 keV and  (d) is after setting $\tau_{\mathrm{cyc}}=0$ in the best-fit model for the \texttt{gabs} at 45 keV. The black points correspond to data while the red line represents the model.}
    \label{fig:unfolded-spectra-residues_ast1} 
\end{figure*}

\begin{figure*}
    \centering
	\includegraphics[scale=0.65, trim={0 3.0cm 0 1.8cm},  angle=-90]{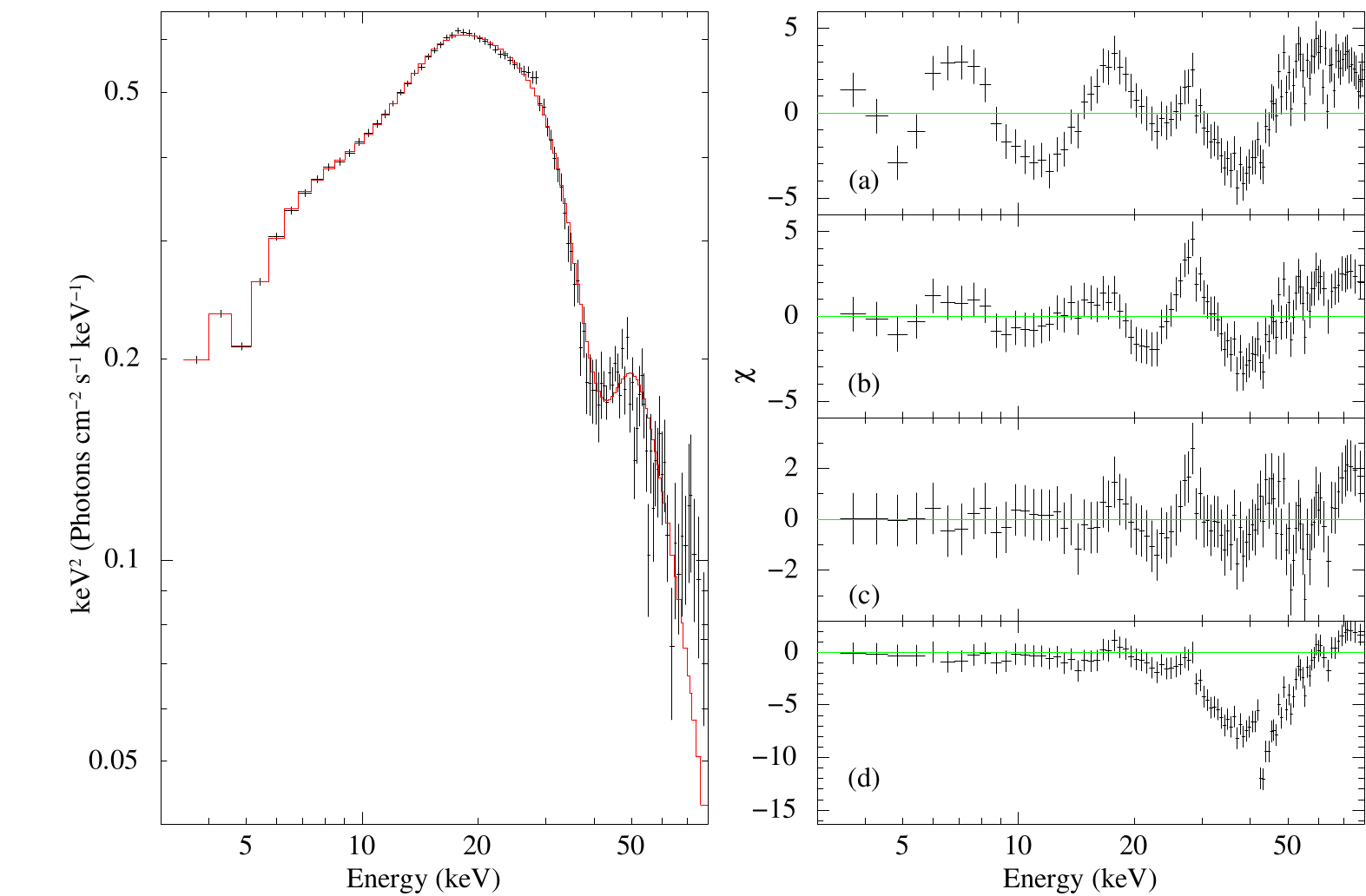}
    \caption{The unfolded spectrum of the \astro\ 2018 observation with the best-fit of \texttt{NPEX} model is shown on the left panel. The right panel shows the residues. Panel (a) is the fit without the cyclotron line at 41 keV and the 10 keV feature (b) is the fit including a \texttt{gabs} at 10 keV (c) is the best fit including a \texttt{gabs} at 41 keV and  (d) is after setting $\tau_{\mathrm{cyc}}=0$ in the best-fit model for the \texttt{gabs} at 41 keV. The black points correspond to data while the red line represents the model.}
    \label{fig:unfolded-spectra-residues_ast2} 
\end{figure*}

\begin{figure*}
    \centering
	\includegraphics[scale=0.65, trim={0 3.0cm 0 1.8cm},  angle=-90]{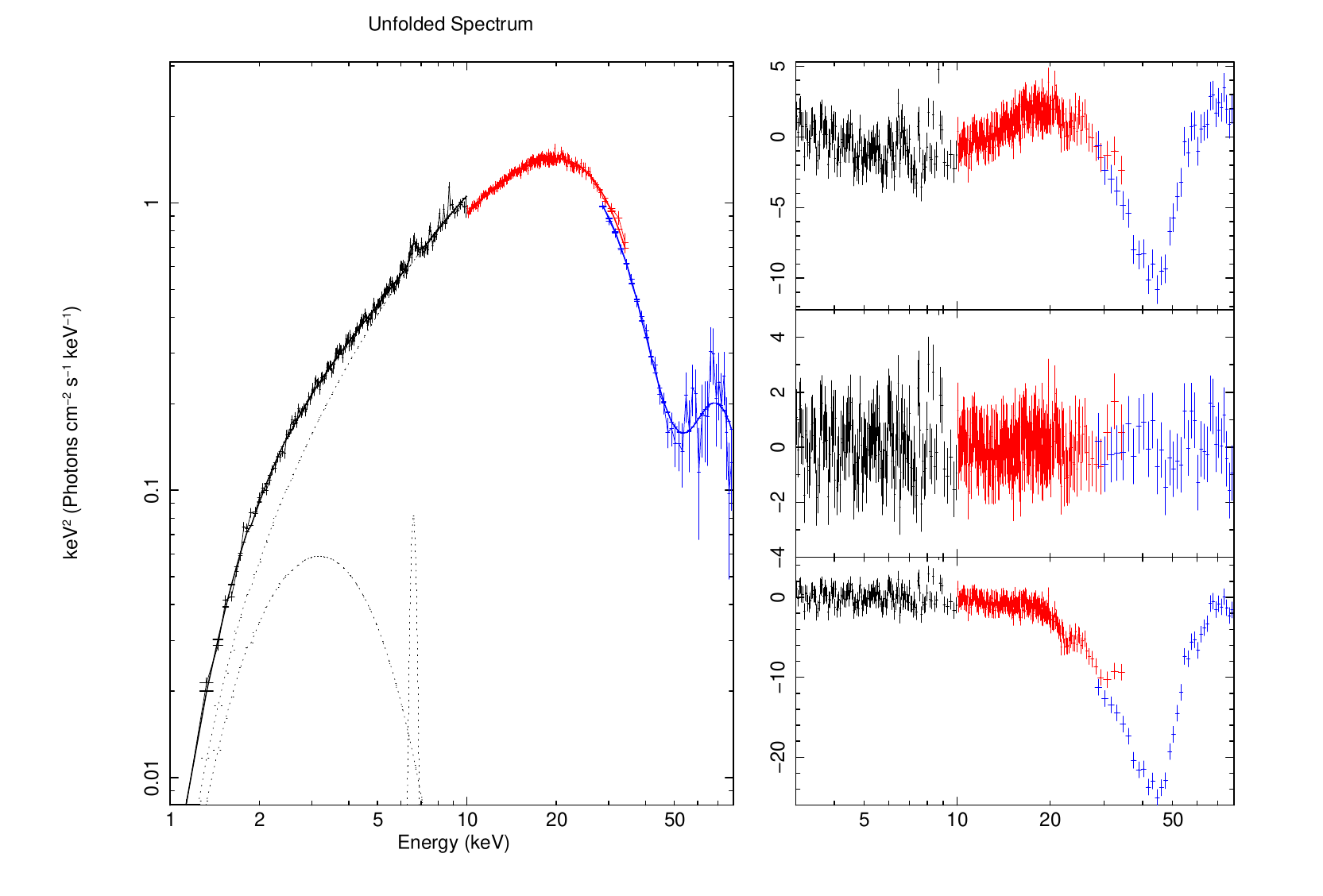}
    \caption{The unfolded spectrum of the HXMT 2018 observation with the best-fit of \texttt{NPEX} model is shown on the left panel. The right panel shows the residues. Panel (a) is the fit without the cyclotron line at 49 keV (b) is the best fit including a \texttt{gabs} at 49 keV and (c) is after setting $\tau_{\mathrm{cyc}}=0$ in the best-fit model for the \texttt{gabs} at 49 keV. The black, red, and blue points correspond to data from LE, ME and HE respectively.}
    \label{fig:unfolded-spectra-residues_hxmt} 
\end{figure*}


\begin{figure}
    \centering
	\includegraphics[scale=0.45, trim={0 3.0cm 0 1.8cm},  angle=-90]{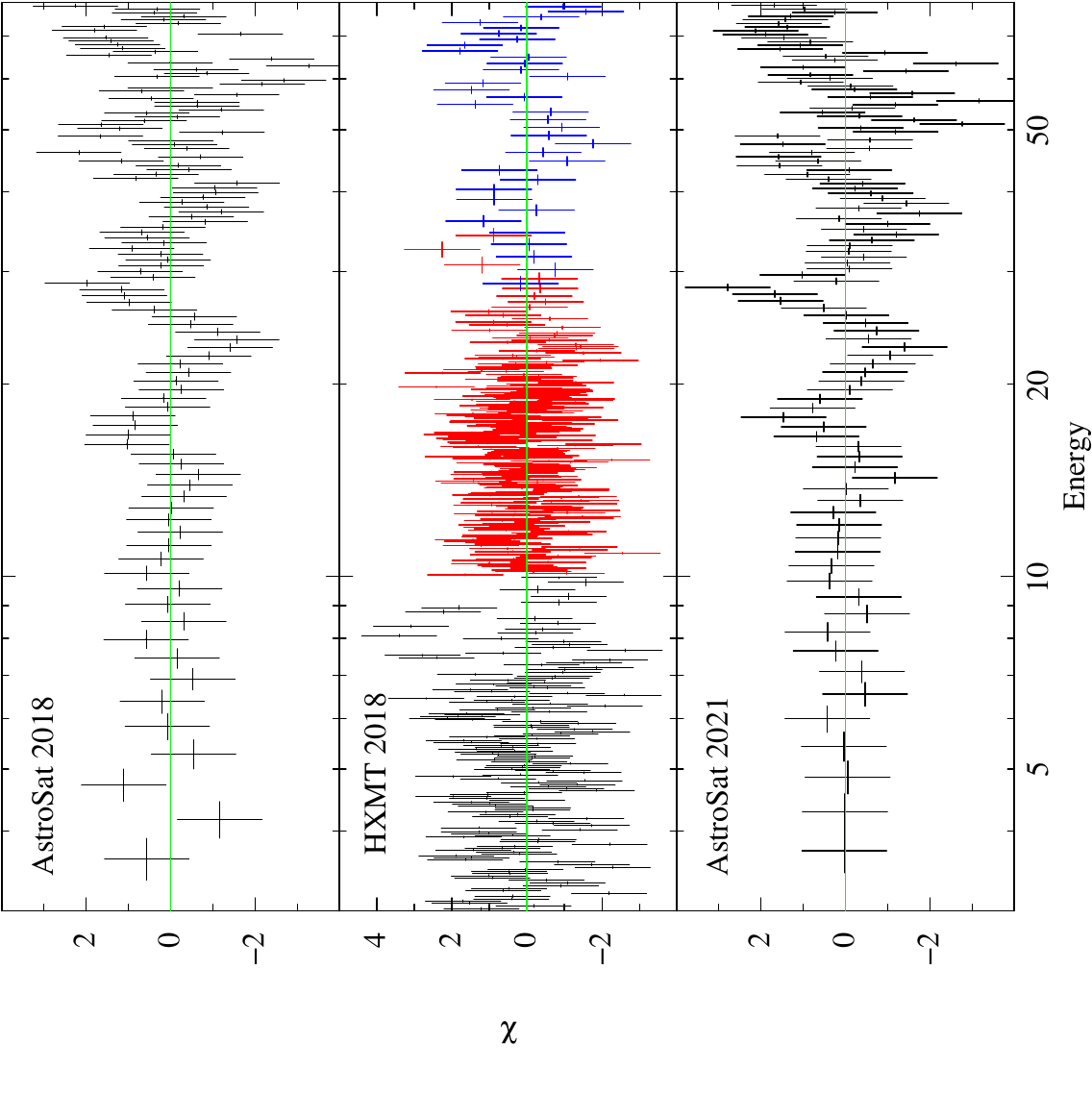}
    \caption{Residuals of the data with the best-fit models. From the top, the panels represent the best-fit residues of the \texttt{CompTT} model for the \astro \ 2018, \hxmt \ 2018, and \astro \ 2021 observations respectively in the 3-79 keV range.}
    \label{fig:compttresid} 
\end{figure}

\subsubsection{Phase-average spectral analysis}

 For both the \astro \ observations, we extracted the source and background spectra using the \texttt{laxpc\_make\_spectra} and \texttt{laxpc\_make\_backspectra} tools for all the layers which accounted for both single and double events. We applied a systematic error of 1 \% between 3 and 10 keV and 0.5 \% between 10 and 80 keV similar to the method used by \citet{Varun_2019a} using the tool \texttt{GRPPHA} unlike the 2 \% over the entire range as used by \citet{chandra_2023} since the systematic errors are more prominent in the lower energies than in higher ranges. We include an edge function fixed at 4.7 keV to account for the Xe-L edge for both observations similar to \citep{Chandra_2020}. 
 
 In the case of the \hxmt \ observation, the energy bands considered for the spectral analysis are 1-10 keV, 10-30 keV, and 25- 80 keV for the LE, ME and HE telescopes respectively. The instrumental background light curves and spectra were generated using the tools provided by the \textit{Insight}-HXMT team: \texttt{lebkgmap, mebkgmap,} and \texttt{hebkgmap}, version 2.05 based on the standard \textit{Insight}- HXMT background models \citep{liao_2020}. To improve the statistics of the spectra, we combined the spectral files from all the exposures using the python script \texttt{combine\_HXMT\_spec.py}\footnote{\url{https://code.ihep.ac.cn/jldirac/insight-hxmt-code-collection/blob/master/version2.04/combine_HXMT_spec.py}}. While performing the simultaneous spectral fitting of the \hxmt \ data from LE, ME and HE detectors, we introduced a relative normalization of LE and HE with respect to the ME. The best spectral fit was obtained for a normalization factor of 1.15 for LE and 0.91 for HE.

 \textit{AstroSat 2018 and 2021}: The commonly used continuum models such as \texttt{HighEC}, \texttt{FDcut} \citep{Tanaka_1986}, \texttt{NPEX} \citep{Makishima_1999}, and \texttt{CompTT} \citep{Titrachuk_1994}, all were found to fit the spectral continuum quite we well as also noted by \citet{chandra_2023}. In particular, the \texttt{NPEX} and \texttt{CompTT} models resulted in very similar fits.   However, for the sake of brevity, we present the fits only from the \texttt{NPEX} and \texttt{CompTT} models. The functional form of the \texttt{NPEX} model used here is as defined in \citet{Devaraj_2022}.  Unlike the approach taken in \citet{chandra_2023}, where data from \textit{AstroSat}/SXT was also utilized, we focused solely on the \textit{AstroSat}/LAXPC data because of SXT readout time of $\sim 2.4$ s which makes pulse phase-resolved spectroscopy for XTE J1946+274 difficult with this instrument. Since \textit{AstroSat}/LAXPC has only coverage above, 3 keV, constraining the N$_\mathrm{H}$ well is not possible. Therefore we fix the value of the \texttt{TBabs} component which accounts for the soft X-ray interstellar absorption to $1.01 \times 10^{22}$ cm$^{-2}$ \citep{HI4PI}. For the \texttt{NPEX} model we obtain the index $\alpha_{\mathrm{NPEX}}\sim$0.36 and $\sim 0.6$ with an e-fold $\sim 8.2$ and $\sim12.2$ keV for the two observations, respectively. For the \texttt{CompTT} model, we obtain a seed photon temperature, T$_0 \sim 1.5$ and $\sim 1.48$ keV and electron temperature kT$_e$ $\sim 8.6$ and $\sim 8.84$ keV for the two observations, respectively. In addition to these continua, we were required to introduce a Gaussian absorption feature around 10 keV to account for the absorption-like residuals, as can be seen from the right panel (a) of Fig. \ref{fig:unfolded-spectra-residues_ast1} and of Fig.\ref{fig:unfolded-spectra-residues_ast2}. This feature, also known as the "10 keV" feature, has been seen in several X-ray pulsars and appears in the spectrum of XTE J1946+274 independent of the flux, pulse phase and the instrument used to make the observation\citep{Coburn_2002, hemanth_2023} and has been attributed as an artefact of modelling of the continuum of these pulsars. In some cases, based on the choice of certain model components such as using a higher blackbody temperature component as used by \citet{Gorban_2021}, the feature can be adjusted. However, as we can see in panel Fig. 7(c) and Fig. 8(c) of \citet{chandra_2023}, there is still a presence of absorption-like features around 10 keV even after the best-fit. After modelling the continuum, we can see the presence of the absorption feature around 44 keV in the right panel (b) of Fig. \ref{fig:unfolded-spectra-residues_ast1} and Fig. \ref{fig:unfolded-spectra-residues_ast2}. This feature, which is interpreted as a cyclotron absorption was fitted with a Gaussian absorption feature (\texttt{gabs}) resulting in a line centroid energy of 44.6 keV for the 2018 observation and $\sim41$ keV for the 2021 observation. The widths of both these features are less than 6 keV and the lines have an optical depth between $0.6-0.7$. The inclusion of the \texttt{gabs} to account for this feature improved the $\chi^2$ by over 177 for a change in 3 d.o.f resulting in the final $\chi^2 <$ 115 for around 99 d.o.f. Panel (c) of Fig. \ref{fig:unfolded-spectra-residues_ast1} and Fig. \ref{fig:unfolded-spectra-residues_ast2} show the residuals after the best-fit and panel (d) in both figures show the residuals when the optical depth of the cyclotron line in the best-fit is set to 0.  
 
\textit{HXMT 2018:} We modelled the \hxmt \ spectra using the \texttt{NPEX} and \texttt{CompTT} continua to obtain satisfactory fits. Since \hxmt \ has low energy coverage, we were able to constrain the N$_{\mathrm{H}} \sim 1.0$. We require a soft blackbody component with a temperature of $\sim 0.75$ keV similar to the \nustar \ case \citep{Devaraj_2022}. We also added a Gaussian feature at 6.6 keV to account for the Fe emission line. We were unable to probe for the 10 keV feature in this observation because of a lack of significant overlap in  the energy coverage between  the LE and ME detectors around 10 keV. The large cross-normalization factor of 15\% between LE and ME may be due to the presence of this feature. After modelling the continuum, we found significant residuals at around 50 keV which we modelled using a \texttt{gabs}. See right panel (a) and (b) of Fig. \ref{fig:unfolded-spectra-residues_hxmt}. The inclusion of the \texttt{gabs} resulted in a $\Delta \chi^2$>180  and 573 for change in 3 d.o.f for the \texttt{NPEX} and \texttt{CompTT} models, respectively. The final reduced $\chi^2$ was $\sim 1$ for 1511 d.o.f. See Table. \ref{table:Bestfit-table-spec} for best-fit parameters. The cyclotron line is found at a much higher energy at $\sim 50$ keV with a large width of $\sim 11-14$ keV and a high optical depth >1.4. This is the highest line energy and optical depth measured for this source as of yet. 

The bestfit residuals for the three observations using the \texttt{CompTT} model are presented in Fig. \ref{fig:compttresid} and are similar to the fits using the \texttt{NPEX} model.


\subsubsection{Phase-resolved spectral analysis}
\begin{figure*}
    \centering
	\includegraphics[scale=0.35,trim={0 0.5cm 0 0.3cm}]{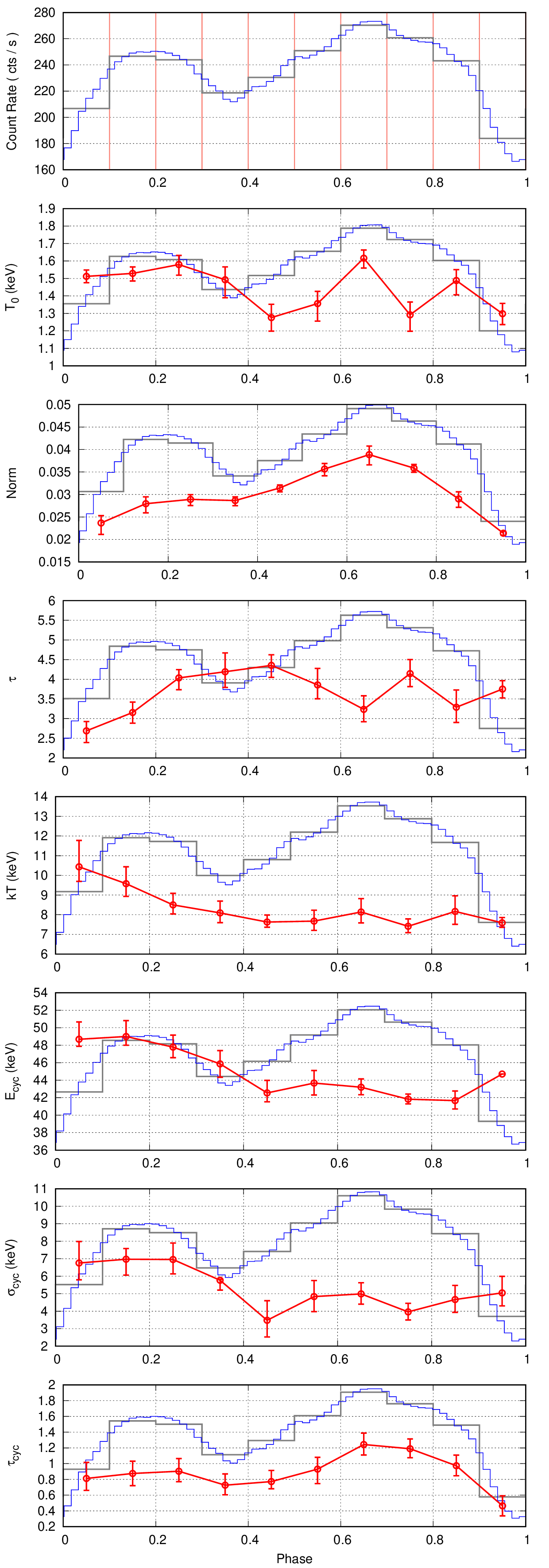}~~
 \includegraphics[scale=0.35,trim={0 0.5cm 0 0.3cm}]{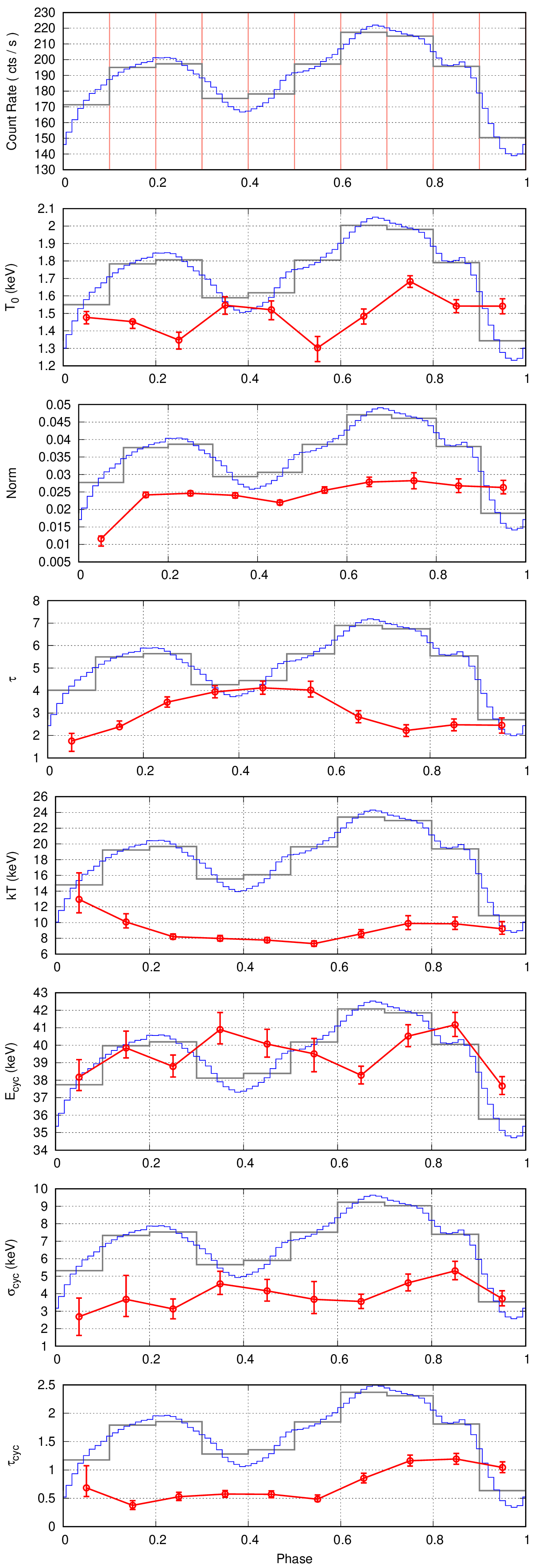}
	    \caption{\textbf{\astro\ 2018 }(left) and \textbf{\astro\ 2021 }(right) \textbf{:}  Variations of the spectral parameters with the pulse phase for \texttt{CompTT} model. 1 $\sigma$ errors are shown for all the parameters. The topmost panel represents the pulse profile (blue) i.e. the count rate with phase. The subsequent panels correspond to the seed photon temperature T$_{0}$, norm, optical depth $\tau$, plasma temperature $kT$, \ecyc, $\sigma_{\texttt{cyc}}$ and $\tau_{\texttt{cyc}}$ respectively. The grey step curve represents the phase bin size as well as the count rate in each phase bin. The two curves have been re-scaled and laid in the background of all the subsequent panels as a reference to see the variation of the spectral parameters with the changes in the pulse profile. }
    \label{fig:phase-resolved-ast_comptt}
\end{figure*}

The evolution of the pulse profiles with energy in Fig. \ref{fig:BBpulseprofiles-all_ast1} is indicative of the evolution of the spectral parameters with the pulse phase. Previously, there were reports of the line being present in only some of the phases using data from \beppo \ \citep{Doroshenko_2017} and this was found to be the case with data from \nustar \ as well where the line was not very significantly detected in the first of the two peaks in the double-peaked structure \citep{Devaraj_2022}. The line energy was also found to remain nearly constant with the pulse phase. To probe this further at different luminosities of the source, we performed phase-resolved spectroscopy with the \astro \ and \hxmt \ data. 
For the \astro \ data, using the appropriate period and period derivatives, we extracted the spectrum in 10 equally spaced bins. We present the phase-resolved spectroscopy results using the \texttt{CompTT} model here since it is a more physical model than \texttt{NPEX}. Fig. \ref{fig:phase-resolved-ast_comptt} shows the variation of the parameters of the \texttt{CompTT} model with the pulse phase for 2018  and 2021 \astro \ observations. While fitting the 2018 \astro\  observation, the $\chi^2$ for each phase bin was around 115 for 99 d.o.f. Except for the phase bin 0.9--1.0 in Fig. \ref{fig:phase-resolved-ast_comptt} (left panel), an improvement in $\chi^2$ of over 70  was observed for every phase bin when including a \texttt{gabs} feature for a change in 3 d.o.f around the cyclotron line energy. The largest change of $\Delta\chi^2 \sim$ 167 was observed for phase 0.7--0.8. For the analysis in the 0.9--1.0 phase bin, we fixed the line energy to the phase average value and allowed the width and optical depth to vary freely. For the 2021 \astro \ observation, we found the line to be clearly detected in all the phases with the change in $\Delta \chi^2$ > 80 except for the 0.0-0.1 phase bin where the improvement was only around 30 for a change in 3 d.o.f when including a \texttt{gabs} feature (see right panel of Fig. \ref{fig:phase-resolved-ast_comptt}). The average $\chi^2$ of the best-fit of each phase bin was $\sim100$ for 96 d.o.f. The line was most strongly detected in the 0.9--1.0 phase bin as can also be seen from its relatively high optical depth (see right panel of Fig. \ref{fig:phase-resolved-ast_comptt}).

The pulse phase dependence of the spectral continuum parameters seems similar between the two \astro \ observations in 2018 and 2021. The same trend can also be seen with the results from \nustar \ (see Fig. 6 of \citealt{Devaraj_2022}). However, the difference arises because the cyclotron line is detected in almost all the phases in the \astro \ observations and shows significant variation with the pulse phase in the 2018 observation with \astro. 
The line energy varies between 42 keV and 50 keV for the 2018 observation while exhibiting a decreasing trend when moving from the first peak to the second peak (see right panel of Fig. \ref{fig:phase-resolved-ast_comptt}). However, the line's optical depth is higher, with a smaller width in the second peak. For the 2021 observation, though the variation of the line energy with the pulse phase is small, the optical depth of the line is higher in the second peak as compared to the first.

\begin{figure}
    \centering
	  \includegraphics[width=\columnwidth]{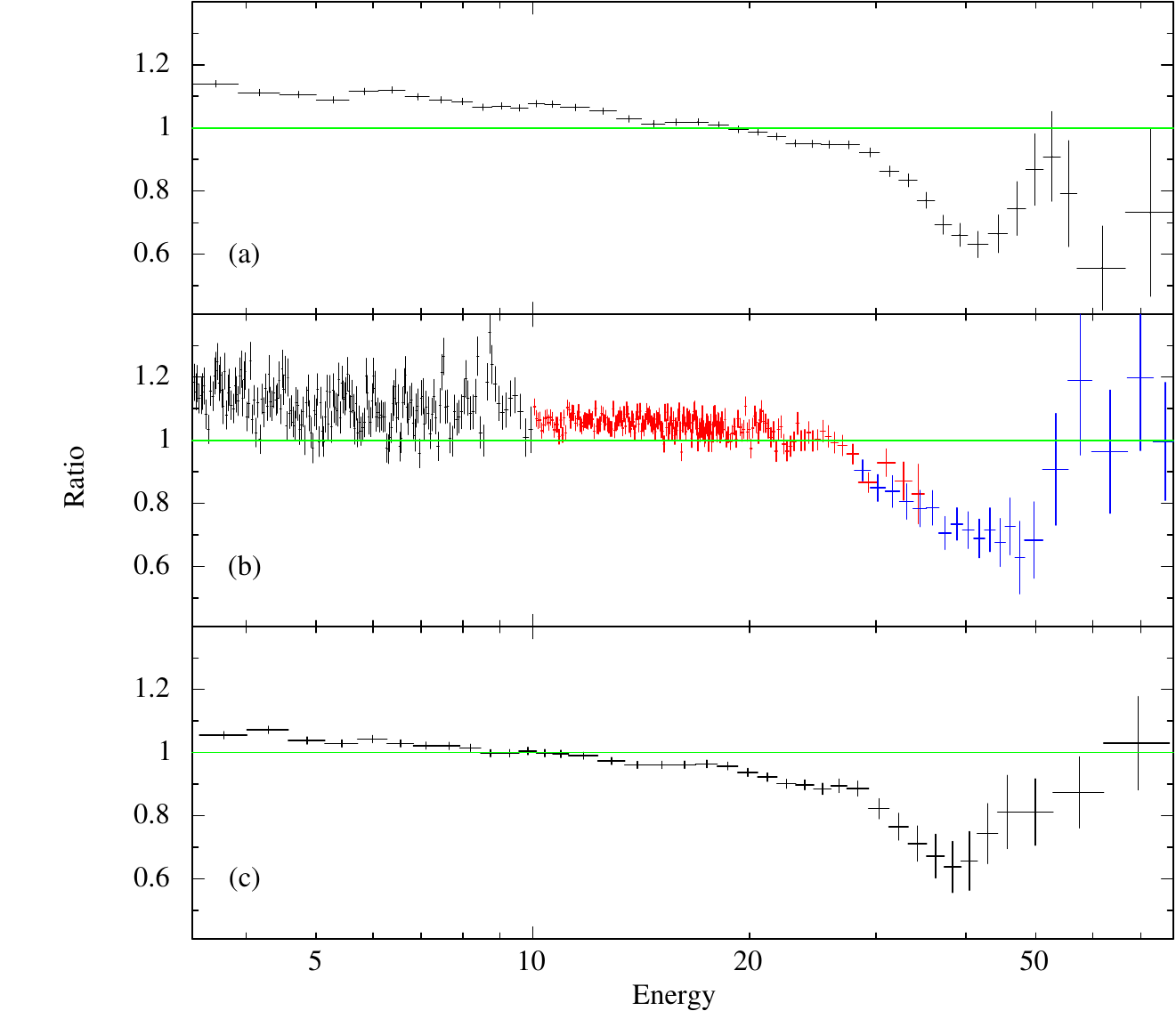}
	    \caption{The ratio of the data in the phases corresponding to the second peak with the best-fit model of the spectra of the phases corresponding to the first peak; indicating a clear variation of the cyclotron line depth with the rotational phase. Panel (a) 2018 \astro\ observation, (b) 2018 HXMT observation and (c)  2021 \astro\ observation.}
	    \label{fig:ratioplots}
\end{figure}

Due to limited statistics, we performed phase-resolved spectroscopy on \hxmt \ observation only in 2 phase bins, corresponding to the first peak and second peak. The presence of the line in the first peak was not very significant. The inclusion of a \texttt{gabs} feature resulted  in an improvement in $\chi^2$ of only 70. The best-fit without the line had a $\chi^2$ of 1618 for 1511 d.o.f. The line was, however, firmly constrained in the second peak with \ecyc$\sim 47.8$ keV with a width of  $\sim$13 keV and optical depth of $\sim$ 1.9. We then produced Fig. \ref{fig:ratioplots}, where the plots represent the ratio of the data in the phases corresponding to the second peak with the best-fit model of the spectra from the phases corresponding to the first peak. From this ratio, we can see that in every observation, similar to the nature of the two peaks as observed with \nustar, the presence of the cyclotron line is much stronger in the second peak than in the first peak.


\begin{table*}
 \bgroup
 \def\arraystretch{1.9}
 \caption{Best-fitting phase-averaged spectral parameters. Errors are reported at 90 \% confidence. }
\begin{center}

    \begin{tabular}{|l|l|l|l|l|l|l|}
        \hline 
 \hline

    ~ Parameters~ & \multicolumn{2}{c|}{~Astrosat I~} &\multicolumn{2}{c|}{~HXMT~} & \multicolumn{2}{c|}{~Astrosat II ~}\\
        \hline 

        &~ NPEX~&~CompTT~ & ~NPEX~ & ~CompTT~ & ~NPEX~ & ~CompTT~ \\
        \hline
        \hline
        ~N$_\mathrm{H}$[$\times 10^{22}cm^{-2}$]~  &    ~1.01\,(fixed)~   &  ~1.01\,(fixed)~   &  ~$1.03_{-0.09}^{+0.10}$~  & ~$0.901_{-0.074}^{+0.079}$~        &  ~1.01\,(fixed)~                 & ~1.01\,(fixed)~ \\
     ~$\alpha_{\mathrm{NPEX}}$~  &   ~$0.358_{-0.031}^{+0.037}$~    &   ~\ldots~   &       ~$<0.15$~           & ~\ldots~                      &  ~$0.606_{-0.091}^{+0.068}$~  & ~\ldots~                 \\
      ~f$_{\mathrm{NPEX}}[\times 10^{-3}]$~ &        ~$0.43_{-0.14}^{+0.17}$~ &    ~\ldots~   &    ~$0.15_{-0.15}^{+1.4}$~   &  ~\ldots~                      & ~$<0.02 $~ & ~\ldots~                \\
    ~E$_{\mathrm{fold}}$ kT (keV)  ~    &   ~$8.16_{-0.49}^{+0.62}$~           &     ~\ldots~    &  ~$10.8_{-2.9}^{+3.6}$~          &  ~\ldots~                       & ~$12.2_{-1.5}^{+1.2}$~        & ~\ldots~ \\
     ~Continuum Norm$[\times 10^{-2}]$~    &   ~$4.91_{-0.24}^{+0.25}$~    &   ~$2.92_{-0.18}^{+0.16}$~     &     ~$2.32_{-0.10}^{+0.56}$~    & ~$3.39_{-0.17}^{+0.18}$~  & ~$4.77_{-0.39}^{+0.39}$~ &  ~$2.36_{-0.16}^{+0.19}$~  \\
      ~CompTT $\mathrm{T}_0$  (keV)   ~   &      ~\ldots~                       &    ~$1.467_{-0.072}^{+0.066}$~    &    ~\ldots ~                          &  ~$1.81_{-0.14}^{+0.17}$~            &  ~\ldots ~                      &  ~$1.49_{-0.06}^{+0.06}$~      \\
      ~CompTT kT    (keV) ~ &       ~\ldots ~                    &     ~$8.56_{-0.53}^{+0.65}$~      &       ~\ldots~                    & ~$8.5_{-0.4}^{+0.5}$~              & ~\ldots   ~                     & ~$8.84_{-0.48}^{+0.61}$~\\  
      ~CompTT $\tau$~ & ~\ldots~    &      ~$3.52_{-0.31}^{+0.33}$~       &      ~ \ldots~                      &  ~$5.5_{-0.32}^{+0.31}$~           & ~\ldots     ~                   & ~$2.99_{-0.31}^{+0.29}$        ~ \\

~Blackbody kT (keV)  ~ & ~\ldots    ~                       &   ~\ldots  ~                      &    ~$0.74_{-0.06}^{+0.04}$  ~       &  ~$0.78_{-0.05}^{+0.05}$   ~      &  ~\ldots      ~                  &  ~\ldots~ \\
~Blackbody Norm~  &  ~\ldots    ~                       &     ~\ldots  ~                      &     ~$40_{-7.5}^{+9.5}$  ~              &  ~$64_{-10.9}^{+14.5}$      ~        &  ~\ldots    ~                    &  ~\ldots~\\
~E$_{10\mathrm{ keV}}$   (keV) ~ & ~$10.7_{-0.2}^{+0.2}$     ~         &      ~$9.41_{-0.46}^{+0.34}$  ~       &     ~\ldots     ~                      &  ~\ldots  ~                         & ~$10.87_{-0.21}^{+0.19}$  ~      &  ~$8.92_{-0.58}^{+0.40}$ ~     \\
~$\sigma_{10\mathrm{ keV}}$   (keV)~ & ~$2.24_{-0.30}^{+0.32}$  ~           &     ~$3.1_{-0.41}^{+0.47}$  ~        &       ~\ldots     ~                      & ~\ldots   ~                        & ~$2.71_{-0.34}^{+0.37}$ ~        &  ~$3.8_{-0.46}^{+0.55}$~  \\
~$\tau_{10\mathrm{ keV} }$ ~ & ~$0.146_{-0.018}^{+0.018}$ ~       &      ~$0.276_{-0.056}^{+0.062}$  ~   &       ~\ldots     ~                      &  ~\ldots    ~                       & ~$0.173_{-0.024}^{+0.025}$~     &  ~$0.42_{-0.09}^{+0.11}$~ \\
~E$_{\mathrm{Fe}}$   (keV)~  & ~\ldots  ~                         &      ~\ldots  ~                      &       ~$6.603_{-0.027}^{+0.054}$  ~       &  ~$6.603_{-0.053}^{+0.052}$  ~       & ~\ldots ~                       &  ~\ldots~ \\
~$\sigma_{\mathrm{Fe}}$    (keV)~ &~\ldots   ~                        &    ~\ldots                    ~    & ~$0.141_{-0.054}^{+0.069}$  ~      &  ~$0.138_{-0.053}^{+0.065}$ ~       &  ~\ldots  ~                      &  ~\ldots ~\\
~Norm$_{\mathrm{Fe}}$[$ \times 10 ^ {-3}$]~ & ~\ldots ~                          &    ~\ldots  ~                      &    ~$0.59_{-0.16}^{+0.18}$ ~ & ~$0.59_{-0.16}^{+0.18}$~  & ~\ldots  ~                      & ~ \ldots  ~\\
~E$_{\texttt{cyc}}$   (keV) ~    &        ~$44.6_{-1.3}^{+0.8}$~            &      ~$44.7_{-1.3}^{+1.4}$~           &       ~$49.0_{-1.4}^{+1.8}$~                &  ~$49.2_{-1.6}^{+2.0}$~           &  ~$40.4_{-0.9}^{+1.2}$~         &  ~$40.8_{-1.1}^{+1.4}$~         \\
~$\sigma_{\texttt{cyc}}$   (keV)  ~  &      ~$5.9_{-1.0}^{+1.1}$  ~               &      ~$6.0_{-1.1}^{+1.2}$  ~            &    ~$11.8_{-1.1}^{+1.3}$ ~             &  ~$14.4_{-1.4}^{+1.8}$  ~            &  ~$4.9_{-0.9}^{+1.1}$~          & ~$5.4_{-1.0}^{+1.4}$ ~             \\
~$\tau_{\texttt{cyc}}$~  &     ~$0.72_{-0.09}^{+0.11}$  ~       &    ~$0.73_{-0.10}^{+0.12}$  ~    &       ~$1.43_{-0.17}^{+0.27}$    ~        & ~$1.86_{-0.24}^{+0.33}$   ~         &  ~$0.579_{-0.061}^{+0.064}$   ~   &  ~$0.626_{-0.062}^{+0.068}$ ~     \\
\hline
~Flux   (3-79 keV)[10$^{-9}$ erg cm$^{-2}$ s$^{-1}$]~	&  $2.55_{-0.03}^{+0.03}$     &     $2.55_{-0.03}^{+0.03}$   &  $4.22_{-0.02}^{+0.02}$       &$4.17_{-0.02}^{+0.02}$&$1.78_{-0.03}^{+0.03}$  & $1.78_{-0.03}^{+0.03}$	   \\
~Tot-Chi-sq(d.o.f)[w/o-cyc]~  	               	& 388(102)   &   386(101)                    &  1948(1511)&  1700(1511)       &  386(101)& 388(99)       \\

~Tot-Chi-sq(d.o.f)[w/-cyc]~  	            	&112(99)                        & 109(98)  & 1375(1508)&              1522(1508)         &                110(98)        &                  114(96)            \\

         \hline
    \end{tabular}

 \end{center}
  \label{table:Bestfit-table-spec}

 \egroup
\end{table*}


\par


 \begin{figure}
    \centering
	\includegraphics[scale=0.37, trim={0 3.0cm 0 1.8cm},  angle=-90]{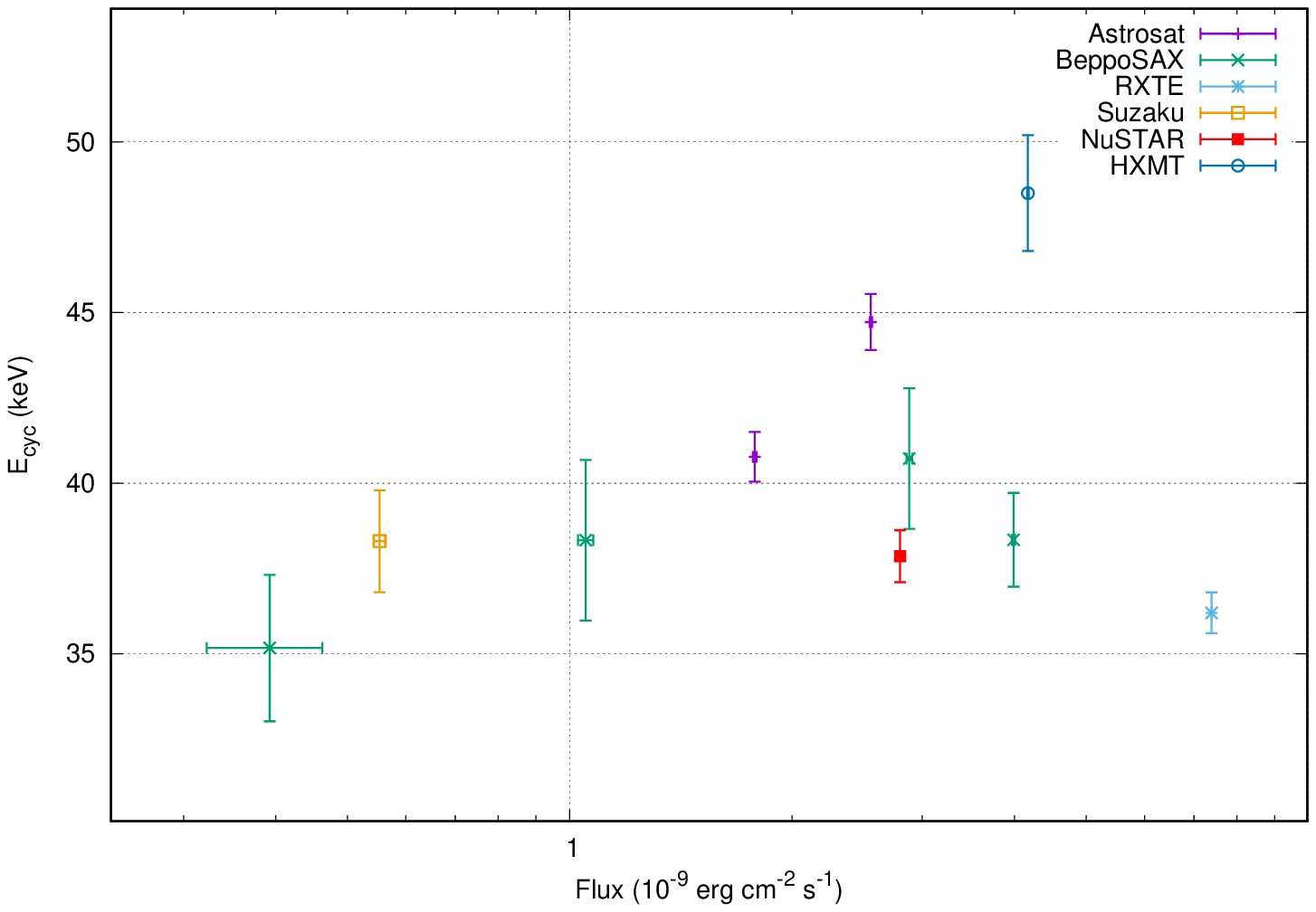}
    \caption{Plot of E$_{\texttt{cyc}}$ and flux. Note that the flux axis is presented using a log scale.}
    \label{fig:EcycVSFlux} 
\end{figure}


\section{Discussion}

The presence of a 37 keV cyclotron line in XTE J1946+274 has been a matter of some debate since its discovery in 1998 by \citet{Heindl2001}. In the \textit{RXTE}/PCA data of another outburst in 2010, \citet{Caballero_2010} found some residuals around 35 keV but not significant enough to justify an additional absorption model component. Later \citet{Muller2012} reported the presence of a 25 keV cyclotron line with low significance (< 2$\sigma$) using \textit{RXTE} observations of the 2010 outburst. The results from the earliest observations using \textit{RXTE} \citep{Heindl2001}, \textit{BeppoSAX} \citep{Doroshenko_2017}, \textit{Suzaku} \citep{Maitra2013,Marcu_Cheatham_2015} and \nustar \ \citep{Gorban_2021, Devaraj_2022} have all consistently suggested the line energy be around 37 keV. However, from the \astro \ and \hxmt \ observations we find that the line energy had significantly changed to about 44 keV during the rising phase of the outburst and to 50 keV at the peak. The line energy during the declining phases of both the 2018 and 2021 outbursts was closer to 40 keV. This is the first time such a change in the line energy has been found for this source. The line energy may vary depending on the luminosity of the source. Sources like Her X-1 \citep{Staubert_2007} and GX 304-1 \citep{klochkov_2012} have exhibited a positive correlation with luminosity while V0332+53 \citep{Vybornov_2018} and A0535+26 \citep{kong_2021} have exhibited a negative correlation with the luminosity. The nature of the correlation and the possibility of a change in the nature of the correlations themselves have been explained using the changing accretion regimes of the neutron star depending on the mass accretion rate. A positive correlation is observed when the neutron star is in the Coulombic shock regime while a negative correlation between \ecyc and luminosity is associated with the radiation shock regime \citep{Becker_2012}. As the mass accretion rate increases the transitions from the Coulombic shock regime to the radiation shock regime occur while crossing the critical luminosity, \Lc \citep{Mushtukov_2015a}. Only two of the sources mentioned above, V0332+53 and A0535+26 have been known to exhibit both a positive and negative correlation. One thing in common with both these sources is that they have been observed over a large range of fluxes.

XTE J1946+274 is also a source that has presented itself over a range of luminosities, however, due to the irregular nature of its outbursts, the number of observations is limited. \citet{chandra_2023} suggested that there may be a positive correlation that switches to a negative correlation above a luminosity of $5 \times 10^{37}$ erg s$^{-1}$ (however their compilation of the list suffers from double-counting such as multiple cyclotron line values from the same observation). In Fig. \ref{fig:EcycVSFlux}, we present a plot of the variation of \ecyc with flux including the results from the \astro \ and \hxmt \ observations. More observations are required to constrain the nature of the correlation. A change in the accretion regime may also manifest itself in the shape of the pulse profiles like in the case of Swift J0243.6+6124 \citep{Wilson-Hodge_2018} and 4U 1626-67 \citep{Beri2014, Sharma2023} when the geometry changes from a 'pencil' beam to a 'fan' beam. However, over the course of the outbursts, the pulse profiles retain their general structure, suggesting that there probably is no change in the accretion geometry of XTE J1946+274.

Pulse phase-resolved spectroscopy is the best tool at our disposal to trace the accretion geometry of the neutron star. Using \textit{Suzaku} observation of the 2010 outburst \citet{Maitra2013} reported a 36\% variation in the \ecyc\ with the pulse phase with the line energy being higher in the first peak as compared to the second peak and the depth of the line to be higher in the second peak. The 2018 \astro \ observation shows a similar trend, with the line's centroid having a higher energy and low optical depth in the first peak and a low energy line with a high depth in the second peak. However, the 2021 \astro \ observation does not show a clear trend in the variation of the line energy with the pulse phase. From the \nustar \ observation of this source, it was found that the presence of the cyclotron line in the first peak was not significant \citep{Devaraj_2022}. Even in the \hxmt \ data, the presence of a line was not significant in the first peak. Similar behaviour of the line was also reported by \citet{Doroshenko_2017} using the brightest \beppo \ observation of the 1998 outburst, where they did not find a significant presence of the line in the first peak. However, the one thing common in all the cases is the higher strength of the cyclotron line in the second peak as compared to the first, which can be clearly seen from Fig. \ref{fig:ratioplots} and Fig. 7 from \citet{Devaraj_2022}. The ratio plots and the pulse profiles (Fig. \ref{fig:BBpulseprofiles-all_ast1}) also show that the first peak is harder than the second peak.

The maxima of the pulse profiles being separated by a phase of almost 0.5 and the unchanging structures of the pulse profiles, irrespective of the luminosity, indicate that two peaks are likely from the two poles of the neutron star. The line energy in the first peak showing a significant variation with luminosity, while the line energy remaining nearly the same in the second peak would indicate that transitions in the accretion regimes occur at different luminosities. This could be explained if we assume that the magnetic field strengths at the two poles were different. The emission from the first pole with a higher field strength would be from an accretion column with a smaller radius, while the second peak's emission would be from an accretion column with a larger radius. The critical luminosity of the two poles would also be different, with the first peak having a lower \Lc and the second peak having a higher \Lc. Transitions in accretion regimes and hence the changes in the \ecyc would occur at lower luminosities at the first peak as compared to the second. 

For the neutron star to have varying polar cap radii or different surface magnetic field strengths at the two poles would be possible only if the structure of the magnetic fields were non-dipolar. Such a configuration may be achieved if there were aligned magnetic dipole and quadrupole moments. As a result of the quadrupole's contribution, the field strength at one pole would be enhanced while the field strength at the other pole would be diminished. Such a configuration has been explored in  \citet{Shakura_1991}, \citet{Long_2007} and \citet{Lockhart_2019} and the pulse profile variations of Her X-1 with phase of the superorbital period have been explained to be from a freely precessing neutron star (NS) with a complex non-dipole magnetic field \citep{Postnov_2013}. We summarise the key points from these works here. Typically, pulsars have been treated as having a dipolar magnetic field configuration. This is a reasonable assumption as the contributions from the higher-order multipole components (quadrupole, octopole, etc.,) drop faster as we move farther away from the neutron star. However, near the neutron star's surface, they may significantly contribute to deciding the structure of the magnetic fields. \citet{Lockhart_2019} show that while a pure dipole configuration results in two polar caps with identical radii, the introduction of an aligned quadrupolar component decreases the size of one pole resulting in a stronger field at this pole while the other pole with the larger radius has a weaker field strength. The hotspot of the pole with the smaller radius is also found to be hotter than the other pole with a weaker field strength. With the increase in the contribution from the quadrupolar component, the hotspot of the pole with the weaker field transforms into a ring-shaped hotspot.

\citet{Long_2007} have explored accretion onto stars with these field configurations for different angles between the magnetic axis and rotation axis of the neutron star. They find that the hotspots of the stars with a significant quadrupolar contribution are generally cooler than in the case of a pure dipolar. This is because the accretion flow that hits the star in the dipole+quadrupole case is not accelerated to the extent of the flow in the pure dipolar case. They also find that the angular momentum transfer from the disk to the star viz the accretion torque on the neutron star is more efficient for a dipolar case as opposed to the dipole+quadrupolar scenario. The accretion torque model proposed by \citet{Ghosh_1978} assumes a dipolar configuration. In the case of a non-dipolar field, the coupling between the magnetosphere of the neutron star and the disk would be different and the estimate of the magnetic field strength may not be accurate. The field strengths measured using the accretion torque model and the value estimated using cyclotron lines have been mentioned in  \citet{Kabiraj_2020}. In some cases, the discrepancy between the two values may be due to possible non-dipolar magnetic field configurations. \citet{Wilson2003} and \citet{Sugizaki_2017} have explored the dependence of the accretion torque on the spin change rate of the neutron star for the case of XTE J1946+274. \citet{chandra_2023} use two decades of \textit{Fermi/GBM} data of this source and study the spin change trend and find that the magnitude of the spin-up rate during the outbursts is over a factor of 2 larger than the spin-down rate during the quiescence.

The drop in the significance of detection of the line in the first peak also occurs at higher luminosities. The observations that show this feature are \nustar, \hxmt \ and \beppo \, which have been observed at higher luminosities than \astro \ and \textit{Suzaku} (See Fig. \ref{fig:EcycVSFlux}). The process of photon spawning may provide a possible explanation for this behaviour. Spawned photons are emitted when electrons that were previously excited to higher Landau levels de-excite. These spawned photons modify the continuum's shape and the depth of the fundamental cyclotron line feature. As transitions to higher harmonics are allowed, the depth of the fundamental line and lower harmonics become shallower \citep{Schonherr2007}. At higher luminosities, it is more likely for the higher harmonics of the cyclotron line to be populated and result in the process of photon spawning. There have been some sources such as KS 1947+300 \citep{furst_2014} and Vela X-1 \citep{Kreykenbohm_2002,Maitra2013} where the fundamental line is weak in most of the phases. Due to the limited sensitivity of the instruments in the higher energy range used in the study so far, we are unable to probe the presence of the harmonic at $\sim 80$ keV in XTE J1946+274.   

\section{Summary}
In this paper, we reported the results of the analysis of \astro \ and \hxmt 's observations of the 2018 outburst and the \astro\ observation of the 2021 outburst. We determined the pulse period of the source after correcting for the orbital motion of the source using all three observations and found that the source is spinning up over the course of the outburst while spinning down in quiescence. We compared the energy-dependent pulse profiles during different phases of the outburst and found that the accretion geometry most likely is unchanging. We modelled the continuum using the \texttt{NPEX} and \texttt{CompTT} and found the line energy to be significantly higher than the previously reported values.

We performed phase-resolved spectroscopy of the three observations and found that the trend of the continuum parameters over the pulse phase and across the outbursts are similar; however, the variation of \ecyc with the pulse phase does not follow the same trend. Even though the line is still significantly detected in all phases of the \astro \ observations, the strength of the line is greater in the second peak compared to the first. Further observations are required to constrain the nature of the correlation between \ecyc and flux. This variation of the line energy observed in the first peak with luminosity but not in the second peak may be because the transitions in accretion regimes may occur at different luminosities at the two poles due to having different surface magnetic field strengths. The low detection significance of the cyclotron line in the first peak at higher luminosities may occur due to the effect of photon spawning.


\section*{Acknowledgements}

We thank Ranjeev Misra for the valuable discussions regarding the reduction of the \astro \ LAXPC data. We thank the referee for the useful comments that improved the quality of this paper. This research made use of data from the \astro\ mission of the Indian Space Research Organisation (ISRO), archived at the Indian Space Science Data Centre (ISSDC). This work has made use of data from the \emph{Insight}-HXMT mission, supported by the China National Space Administration (CNSA) and the Chinese Academy of Sciences (CAS). We  thank the LAXPC Payload Operation Center (POC) and IUCAA science support cell for providing the necessary software tools. This research has also used data from \textit{Swift}-BAT.


\section*{Data Availability}
Data used in this work can be accessed through the Indian Space Science Data Center (ISSDC) at 
\url{https://astrobrowse.issdc.gov.in/astro\_archive/archive/Home.jsp} and \textit{Insight-HXMT} at \url{http://archive.hxmt.cn/}.


\bibliography{bibtex}
\bibliographystyle{mnras}


\bsp	
\label{lastpage}
\end{document}